\documentclass[]{interact}
\usepackage{epstopdf}% To incorporate .eps illustrations using PDFLaTeX, etc.
\usepackage[caption=false]{subfig}% Support for small, `sub' figures and tables

\usepackage[numbers,sort&compress,merge]{natbib}% Citation support using natbib.sty

\usepackage{amsmath}
\usepackage[normalem]{ulem}
\usepackage[colorlinks=true, allcolors=blue]{hyperref}
\usepackage[version=4]{mhchem}

\theoremstyle{plain}% Theorem-like structures provided by amsthm.sty

\theoremstyle{definition}

\theoremstyle{remark}

\newcommand{\be}{\begin{equation}}
\newcommand{\ee}{\end{equation}}

\def\kHz{ \textrm{kHz} }
\def\MHz{ \textrm{MHz} }
\def\GHz{ \textrm{GHz} }
\def\THz{ \textrm{THz} }
\def\Hz{ \textrm{Hz} }
\def\mHz{ \textrm{mHz} }

\newcommand{\dcs}{\ce{D_2CS}}
\newcommand{\hcs}{\ce{H_2CS}}

\begin{document}

\title{Rotational Spectroscopic Characterization of the [D$_2$,C,S] System: An Update from the Laboratory and Theory.}

\author{
\name{Natalia Inostroza-Pino\textsuperscript{a}\thanks{CONTACT N.Inostroza-Pino. Email: natalia.inostroza@uautonoma.cl}, Valerio Lattanzi\textsuperscript{b}\thanks{CONTACT Valerio Lattanzi. Email: lattanzi@mpe.mpg.de}, C. Zachary Palmer\textsuperscript{c}, Ryan C. Fortenberry\textsuperscript{c}, Diego Mardones\textsuperscript{d},Paola Caselli\textsuperscript{b}, Oko E. Godwin \textsuperscript{a},Timothy J. Lee\textsuperscript{e}
 }
\affil{\textsuperscript{a}Universidad Autónoma de Chile, Facultad de Ingeniería, Núcleo de Astroquímica \& Astrofísica, Av. Pedro de Valdivia 425, Providencia, Santiago, Chile; \textsuperscript{b}Center for Astrochemical Studies, Max-Planck-Institut f\"ur extraterrestrische Physik, Gie\ss enbachstr.~1, 85748 Garching, Germany;
\textsuperscript{c}Department of Chemistry \& Biochemistry, University of Mississippi,
University, MS 38655-1848, U.S.A.; 
\textsuperscript{d}Universidad  de Chile, Facultad de Ciencias F\'isicas y Matem\'aticas, Departamento de Astronom\'ia, Camino el observatorio 1515, Las condes, Santiago, Chile; \textsuperscript{e}NASA Ames Research Center, MS 245-3, Moffett Field, CA 94035,
U.S.A.
}}

\maketitle

\begin{abstract}
The synergy between high-resolution rotational spectroscopy and quantum-chemical calculations is essential for exploring future detection of molecules, especially when spectroscopy parameters are not available yet. By using highly correlated \emph{ab initio} quartic force fields (QFFs) from explicitly correlated coupled-cluster theory, a complete set of rotational constants and centrifugal distortion constants for \ce{D2CS} and  \emph{cis}/\emph{trans}-DCSD isomers have been produced. Comparing our new \emph{ab initio} results for {\dcs} with new rotational spectroscopy laboratory data for the same species, the accuracy of the computed \textit{B} and \textit{C} rotational constants is within 0.1\% while the \textit{A} constant is only slightly higher.  Additionally, quantum chemical vibrational frequencies are also provided, and these spectral reference data and new experimental rotational lines will provide additional references for potential observation of these deuterated sulfur species with either ground-based radio telescopes or space-based infrared observatories.

\end{abstract}


\section{Introduction}

The astrochemistry of sulfur-bearing molecules is significantly affected by the elemental depletion of sulfur \citep{laas2019modeling}. Detailed investigation of reactions involving S is necessary to improve the accuracy of chemical models \citep{vidal2017reservoir}.  
Thus, the discovery of new sulfur-bearing species within the interstellar medium (ISM) and a thorough comprehension of their formation pathways and their contributions to sulfur chemistry is essential. As a result, studies of sulfur-containing molecules in interstellar clouds is a dynamically evolving field of research \citep{spezzano2022}. The detection of any species containing element-16 can aid in classifying sulfur chemistry by contrasting the observed abundances with those anticipated by chemical models. 

\ce{H2CS} (thioformaldehyde) is one of the first interstellar, sulfur-bearing  molecular species detected and is abundant in interstellar molecular clouds (e.g. Sagittarius B2 \citep{Sinclair1973}), star-forming regions \cite{1991AA_H2CS}, carbon-rich stars (e.g. IRC+10216 \cite{Agundez2008}), in starburst galaxies \cite{2005_NGC}, and in extra-galactic sources. These detections include also singly and doubly isotopically-substituted species.  For example, in the pre-stellar core L183 three isotopologues, namely \ce{H2C^34S}, HDCS, and {\dcs}, were detected by \citet{Lattanzi2020} along with the main isotopic species {\hcs}.  In addition, carbon-sulfur chemistry involving short and long carbon chains is important in cold dense interstellar environments, as shown by the identification of five new sulfur-containing species, including NCS, HCCS, \ce{H2CCS}, \ce{H2CCCS}, and \ce{C4S} in addition to \ce{C5S} in the cold dark core TMC-1 by \citet{cernicharo2021tmc}.
 
With multiple deuterated molecules, the combined analysis of observations with chemical models is highly effective to trace the molecules across various stages in the formation of stars and planets \cite{Caselli12}. Thus, deuteration plays a pivotal role to gain insights into the intricate nature of interstellar chemistry. 
The first deuterated \ce{H2CS} maps towards the pre-stellar core L1544 were reported by \citet{2022AA_H2CS}. 
\citet{2022AA_H2CS} suggest that the increased deuteration in \ce{H2CS} observed in protostellar cores and comets is probably inherited from the pre-stellar phase. 
Nonetheless, when the chemical models were compared, it became evident that the reaction network responsible for the formation of doubly deuterated \ce{D2CS} is not yet fully understood.

The lack of accuracy in the vibrational frequencies analysis of \ce{H2CS} in the CSO spectral line survey \cite{Schilke2001} led to a thorough theoretical and experimental study by \citet{Mueller+_19} resulting in the determination of the ground state rotational spectroscopic data of seven of these isotopomers including \ce{H2CS}, \ce{HDCS}, \ce{H2C^33S}, \ce{H2C^{34}S}, \ce{H2C^36S}, H$_2^{13}$CS, H$_2^{13}$C$^{34}$S. 
However, the doubly-deuterated \ce{D2CS} species was excluded from the laboratory study, and the analysis by \citet{Mueller+_19} relied on calculations based largely upon a previous quantum chemical calculation by Martin et al. \cite{Martin1994}, old astronomical observations, and low-frequency laboratory data.
The theoretical studies in by \citet{Mueller+_19} work utilized the B3LYP variant of density functional theory \citep{becke98density,lee1988development}, a Møller-Plesset second-order perturbation theory (MP2) \citep{moller1934note} carried out with Gaussian 09 \citep{frisch2009gaussian}, and CCSD(T) \citep{raghavachari1989fifth} calculations using the Mainz Austin-Budapest version of ACESII and its successor CFOUR\footnote{CFOUR, a quantum chemical program package written by J. F. Stanton, J. Gauss, M. E. Harding, P. G. Szalay et al. For the current version, see http://www.cfour.de} employing the cc-pVXZ (X=T, Q, 5) basis set \citep{dunning1989gaussian,dunning2001gaussian} for hydrogen and carbon and cc-pV(X+d)Z basis set for sulfur in order to determine the equilibrium structural parameters of \ce{D2CS}. The anharmonic force field calculations were done to evaluate first-order vibration-rotation parameters \citep{mills19723}. 

Quantum chemistry offers a powerful tool for the comprehensive spectral characterization of sulfur-containing molecules. This is particularly valuable as laboratory-based studies often face challenges such as the absence of definitive rotational and spectroscopic constants, as well as limitations in the accuracy of experimental measurements. Furthermore, symmetry considerations are crucial in this context, as they profoundly influence molecular properties, including geometry, vibrational modes, and various spectroscopic and chemical characteristics. Conducting a precise symmetry analysis, such as determining the molecular point group, can aid in categorizing vibrational modes and predicting their suitability for infrared (IR) spectroscopy. Additionally, molecular symmetry can significantly impact computational efforts, enhancing the accuracy of results obtained through quantum chemistry simulations.
For several years, Professor Timothy J. Lee \footnote{https://orcid.org/0000-0002-2598-2237} has demonstrated the effectiveness of coupled-cluster (CC) theory, a method renowned for its accuracy \cite{Zhao_2014,fortenberry2017detectability,Martin1992,Dateo_tim1994,Martin_tim1995,Martin1994,tim_qff,Huang2009,inostroza2011,Inostroza_2013,Fortenberry2011,Lee1995}. 
This method enables the calculation of a potential energy surface, incorporating advanced corrections such as scalar relativistic corrections, core electron correction, excitation up to triple order, and the use of highly correlated basis sets. These techniques have continued to evolve. Today explicitly correlated CCSD(T)-F12 combined with the previously discussed corrections and basis sets is commonly employed. This approach demonstrates its capability in accurately predicting both vibrational and rotational constants for molecules, with deviations of no more than 7.5 MHz of the corresponding experimental values \citep{watrous_jpca_accuracy,gardner2021highly}.

Furthermore, the astronomical identification of \ce{S2H}, a molecule containing sulfur, for example, was likewise guided by quantum chemical calculations using these methodologies \citep{Zhao_2014,fortenberry2017detectability}.
In addition,
\citet{INOSTROZAPINO2020111273} performed accurate CCSD(T)-F12/cc-pVTZ-F12 calculations of \ce{H2CS} and its isomerized molecules, including HCSH with \textit{cis} and \textit{trans} conformations and 3A$^{\prime\prime}$H$_2$SC. This theoretical work provided rotational and rovibrational spectral data for several isotopic species of \ce{H2CS}, and those results are in excellent agreement with previous quantum chemical data \cite{Galland_H2CS,ochsenfeld1999coupled}. While the \emph{cis}-HCSH isomer exhibits the lowest energy relative to the \ce{H2CS} global minimum, making it the most probable candidate for observation, primarily, in laboratory and/or interstellar medium (ISM) as pointed out earlier by \emph{ab initio} computations, the present study of D$_2$CS still presents observable spectral features.

The growth in theoretical approaches should raise concerns about the state of the \ce{D2CS} molecular constants, especially owing to the additional lack of modern experimental data. Furthermore, the Cologne Database for Molecular Spectroscopy \citep{CDMS} (CDMS\footnote{www.cdms.de}) explicitly states that predictions of {\dcs} above 200 GHz should be viewed with caution, especially if the calculated uncertainties exceed 0.2 MHz. Providing data to compensate for these shortcomings requires highly accurate theoretical and state-of-the-art laboratory investigations of spectroscopic data crucial for understanding the molecular structure and the dynamics of this deuterated variant \ce{D2CS}. This present work focuses on doubly-deuterated isotopologues of \ce{H2CS} such as \ce{D2CS}, \ce{\textit{cis}-DCSD} and \ce{\textit{trans}-DCSD}, and their possible detection in laboratory settings or in the ISM through observations made by observatories like the James Webb Space Telescope. 
In addition, this work compares these computational results with laboratory experimental data obtained for \ce{D2CS}.
The synergy of theory and laboratory work provides essential spectroscopy data of doubly-deuterated species including new data for key conformers that can contribute to astrophysical deuterated molecular detections of sulfur-bearing species.

\section{Laboratory Measurements}

The laboratory experiment was performed with the frequency-modulated free-space absorption cell spectrometer, known as the Center for Astrochemical Studies Absorption Cell (CASAC), developed at our institute \citep{Bizzocchi2017}, and already adopted for the characterisation of other reactive sulfur bearing species \citep[e.g.][]{Prudenzano2018,Lattanzi2018}.

The primary radiation source employed was a frequency synthesizer (Keysight E8257D) precisely synchronised with a 10 MHz rubidium frequency standard (Stanford Research Systems), ensuring utmost accuracy in frequency and phase stabilisation. Subsequently, the radiation generated by the synthesizer was coupled to a Virginia Diodes (VDI) solid-state active multiplier chain, affording exceptional frequency agility and seamless coverage of the 75–1100 GHz frequency range. 
Routing through a 3 meter long and 5 centimetre diameter Pyrex tube, the radiation encountered two hollow stainless steel electrodes, each measuring 10 centimetres in length, and connected to a direct current (DC) power supply of 5\,kW. The discharge region spanning $\sim$\,2 meters was defined by the distance between these electrodes, and could be efficiently cooled by liquid nitrogen. 
Frequency modulation of the radiation was achieved by encoding its signal with a sine-wave at a steady rate of 15 kHz. Upon interaction with the molecular plasma, the signal was detected through the use of a liquid-helium-cooled InSb hot electron Bolometer, (QMC Instruments Ltd.) To derive the absorption signal's second derivative profile (see Figure\,\ref{fig:spectra}), a lock-in amplifier (SR830, Stanford Research Systems) was employed to demodulate the detector output at twice the modulation frequency (\textit{2f} detection). All of this was coordinated and recorded through the computer-controlled acquisition system.

The chemical conditions responsible for producing the doubly-deuterated thioformaldehyde were initially based on those leading to the greatest yield of the main isotopic species {\hcs}. Once the optimization of the latter production was accomplished, the hydrogen sample was substituted with the deuterium for the current experimental phase. Further refinements in the experimental conditions revealed that a 1:6 mixture of CS$_2$ and D$_2$ diluted in an argon buffer gas, proved to be the most efficacious configuration, resulting in a total pressure of 30 mTorr, as measured at the output of the absorption cell. Crucial parameters contributing significantly to signal quality encompassed a standard glow DC discharge at 40\,mA at approximately 0.7\,kV, and the maintenance of a glass wall temperature at approximately 250\,K.\\

\begin{table}
\caption{{\dcs} Spectroscopic parameters}
\label{table:1} \centering
\begin{tabular}{llr@{.}lr@{.}lr@{.}lr}
\hline\hline   
\noalign{\smallskip}
Parameter & unit & \multicolumn{2}{c}{This work} & \multicolumn{2}{c}{\citet{Mueller+_19}}  & \multicolumn{2}{c}{\emph{ab initio}$^a$} & Accuracy$^b$[\%] \\
\hline
\noalign{\smallskip}
   $A-(B+C)/2$      & \MHz  &      132199&508(88)     &     132198&92(26)   &   132093&5    &   0.08   \\
   $(B+C)/2$      & \MHz  &       14200&04229(20)  &     14200&0562(59)  &  14140&1        &   0.4    \\
   $(B-C)/4$      & \MHz  &         352&104853(92)  &      352&10544(15)  &  350&1          &   0.6    \\
   $D_J$    & \kHz  &      12&31836(35)      &     12&49(16)        &     12&081           &   1.9    \\
   $D_{JK}$ & \kHz  &     282&4981(59)       &    290&9(15)         &      280&             &   4.2   \\
   $D_K$    & \MHz  &       5&682(17)        &      5&5             &      5&376           &   5.4    \\
   $d_1$    & \kHz  &      -1&38867(15)      &   -1&4027(23)        &    -1&313            &   5.4    \\
   $d_2$    & \kHz  &      -0&29078(15)      &   -0&28995(37)       &    -0&257            &   11     \\
   $H_{J}$ & \mHz   &       1&27(14)         &   1&3                &      1&322           &   4.1    \\
   $H_{JK}$ & \Hz   &       0&8835(56)       &   0&88               &      0&841           &   4.8    \\
   $H_{KJ}$ & \Hz   &      -4&680(45)        &  -4&7                &     -4&408           &   5.8    \\
   $h_1$    & \mHz  &       2&403(78)        &   3&0                &      2&123           &   12     \\
   $h_2$    & \mHz  &       2&486(40)        &   1&72(11)           &      2&270           &   8.7    \\
   $h_3$    & \mHz  &       0&636(11)        &   0&792(33)          &      0&557           &   12     \\
\hline
\noalign{\smallskip}
$N^c$       &      & \multicolumn{2}{c}{185} \\
\hline
\noalign{\smallskip}
\end{tabular}
\raggedright{Values in parentheses represent 1$\sigma$ uncertainties, expressed in units of the last quoted digit}.\\

\raggedright{$^a$ \citet{INOSTROZAPINO2020111273}.}\\
\raggedright{$^b$ $100 \times (X_{new}-X_{ab initio})/X_{new}$ where $X$ is one of the molecular parameters in column 1.}\\
\raggedright{$^c$ Total number of transitions analysed.}\\

\end{table}

\section{Computational Approach}

Extensive treatment of correlation effects was included using the singles and doubles coupled-cluster method that includes a perturbational estimate of the effects of connected triple excitations, denoted as CCSD(T) within the explicitly correlated F12b formalism, complemented by the corresponding cc-pVTZ-F12 basis set \cite{Dunning1989,raghavachari1989fifth, Dunning1995, Shavitt09, ccreview, Adler07, Knizia09}.  This is often referred to as ``F12-TZ''  \cite{Agbaglo19b, Agbaglo19c}.
The molecular geometry was obtained with the F12-TZ level of theory. Subsequently, displacements about a set of coordinates are performed to generate the energy points required to establish the fourth-order Taylor series expansions of the potential within the intermolecular Hamiltonian, known as quartic force fields (QFFs) \citep{tim_qff}.

The complete Watson \textit{S}-reduced asymmetric top Hamiltonian is adopted, allowing for direct quantitative comparison between the theoretical spectroscopic constants and those from the laboratory.

\section{Discussion}

\subsection{Laboratory Rotational Spectroscopy}
The laboratory search was motivated by the high quality, intensity and production efficiency of {\hcs} appearing in a previous experiment, which also involved a plasma discharge of a CS$_2$, H$_2$, and Ar mixture \citep{Prudenzano2018}. Once the signal of the main isotopic species was optimised, the hydrogen sample was replaced with deuterium. Our search was guided by the lower frequency measurements (up to 58\,GHz \cite{Johnson71,Cox1982}) combined with the data reported from the astronomical observations of \citet{Marcelino2004DeuteratedTI}, as reported in the Cologne Database for Molecular Spectroscopy. A total of more than 150 new rotational transitions with frequencies up to 1.068\THz were recorded in less than a week. 

{\dcs} is a near-prolate asymmetric rotor ($\kappa=-0.98$) with a $C_{2v}$ symmetry and planar structure (inertial defect $\Delta \approx 0.09$ amu$\cdot$\AA$^2$). Due to its symmetry, the dipole moment lies totally on the $a$ inertial axis and its value was determined experimentally by \citet{Cox1982} as $\mu_a = 1.6588$ Debye, about 1\% larger than that of the main isotopic species. Spin-statistics due to the two equivalent D atoms, lead to \textit{ortho} and \textit{para} states with a 2:1 ratio with the \textit{ortho} states described by even values of $K_a$.

\begin{figure*}[htbp]
    \centering
    \includegraphics[scale=0.13]{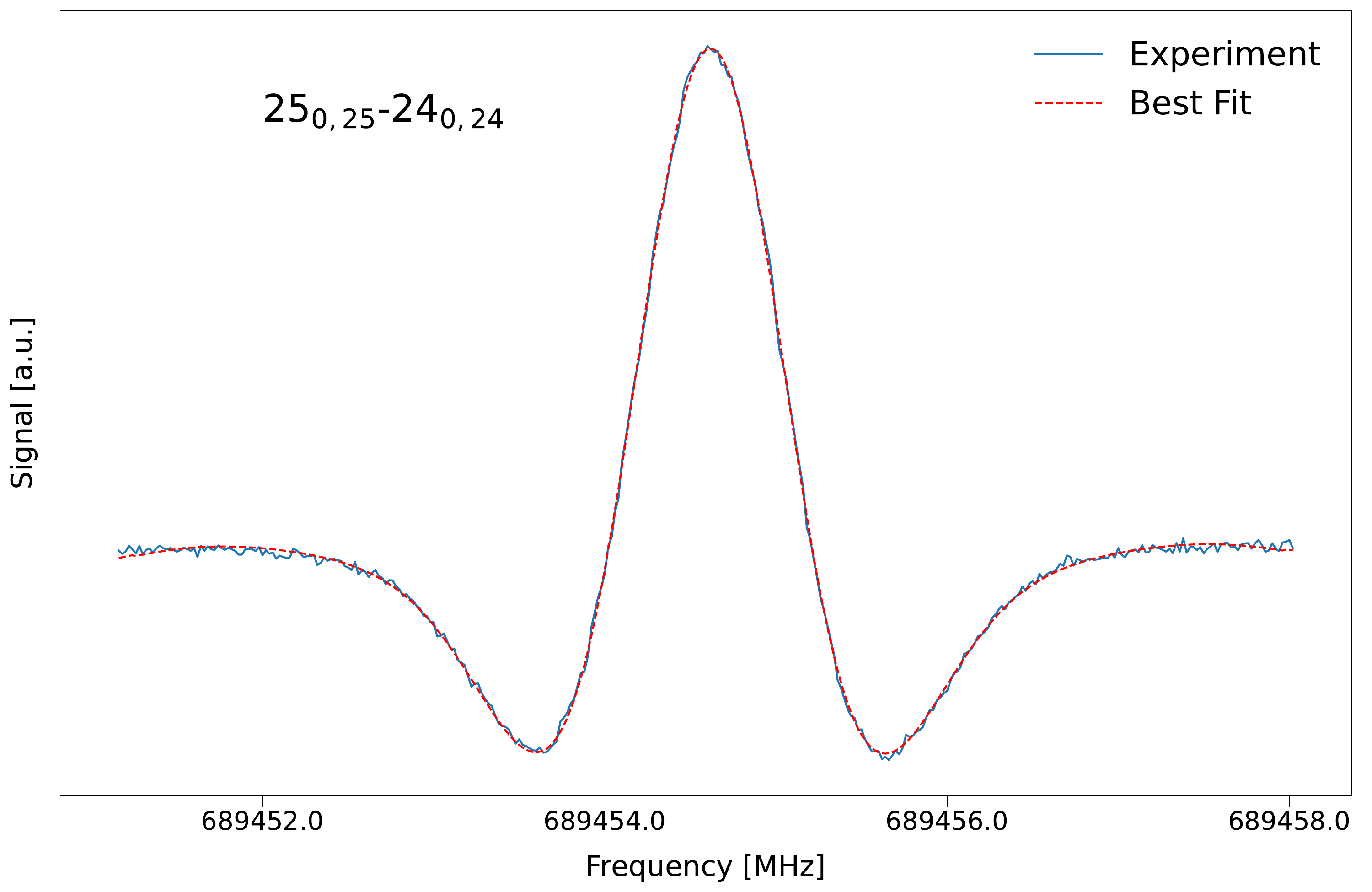}
    \includegraphics[scale=0.13]{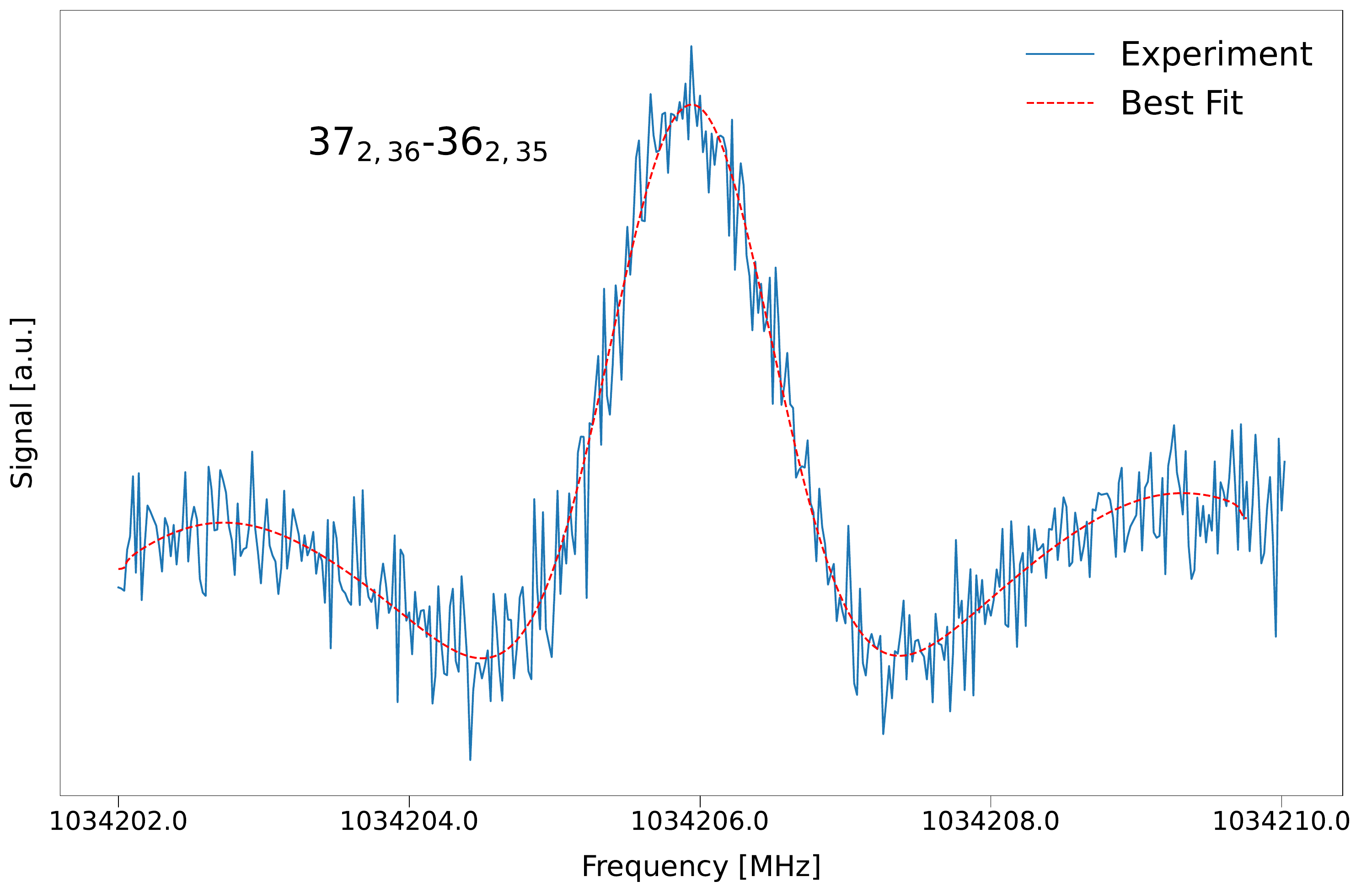}
    \caption{Experimental rotational spectra of \dcs. \textit{a-type} 2\textit{f} absorption features acquired in 73\,s (\textit{left panel}) and 220\,s (\textit{right panel}) integration time, with a 3\,ms time constant. The red dashed lines represent the best fit to a speed-dependent Voigt profile.}
    \label{fig:spectra}
\end{figure*}

To recover the line central frequencies each individual experimental rotational transition was fitted to a modulated Voigt algorithm \citep{Dore2003}, implemented in the data analysis software \textit{QtFit}, part of our in-house python-based libraries for laboratory spectroscopy \textit{pyLabSpec}\footnote{https://laasworld.de/pylabspec.php}; both the complex component of the Fourier-transform of the dipole correlation function (i.e. the dispersion term) and a third-order polynomial were taken into account to model the line asymmetry and a baseline produced by the background standing-waves between non-perfectly transmitting windows of the absorption cell. An example of the quality of this line profile fitting routine is shown in Figure\,\ref{fig:spectra} where a couple of $2f$ absorption features are seen with the modulated Voigt model. From this analysis, and considering the line-width and signal to noise of each rotational transition, we estimated an accuracy of 20-50\,kHz to our experimental data. With the line central frequency in hand, the whole dataset, including the low frequency transitions previously available, was then analysed using the SPFIT/SPCAT suit of programs \citep{Pickett1991} and fitted to a Watson \textit{S}-reduced Hamiltonian for asymmetric-top molecules. In the few cases (in the 3mm wavelength range) where we remeasured a line previously published from astronomical observations \cite{Marcelino2004DeuteratedTI}, the new laboratory frequency was used in the final spectral analysis.

Finally, the frequency range 83 to 1068\,\GHz was covered, recording 185 rotational transitions from 156 independent absorption features ($J_{max}=38, K_{a, max} = 13$), which allowed for the determination of the complete set of quartic and sextic centrifugal distortion constants with a final standard deviation of 21\,kHz. The dimensionless rms obtained is $\sigma_{w} = \sqrt{{\sum_i\left(\delta_i/err_i\right)^2}/{N}}= 0.79$, where the $\delta$ values are the residuals weighted by the experimental uncertainty (\emph{err}) and \emph{N} the total number of transitions analysed.
The parameters were finally re-normalised taking into account the weighted standard deviation of the fit through the PIFORM\footnote{http://info.ifpan.edu.pl/$\sim$kisiel/asym/asym.htm} program and are presented in column 3 of Table\,\ref{table:1}.  The new catalogue obtained from our analysis will be available on the CDMS website after the publication of this work.

\subsection{Ab initio calculations}
 Table \ref{table:1} lists the rotational, quartic, and sextic centrifugal distortion constants of \ce{D2CS} compared with experimental results at the F12-TZ level of theory.  The accuracy between the previously published \emph{ab initio} data \citep{INOSTROZAPINO2020111273} and our experiment for \textit{A, B,} and \textit{C} rotational constants are 0.08\%, 0.4\%, and 0.6\%, respectively. 
 %Major agreements were obtained for distortional constants.

Table \ref{table:2} includes the new data obtained by F12-TcCR QFF. Using these results, the comparison with the experimental constants of \ce{D2CS} are 
\textit{A--(B+C)}/2=132414.6 MHz, \textit{(B+C)}/2=14206.5 MHz; and \textit{(B--C)}/4=352.4 MHz. 
This represents an improvement with respect to experimental values of 0.163\%, 0.045\%, and 0.084\%, respectively.
Based on these benchmarks, the F12-TcCR results should be providing the most reliable \emph{ab initio} distortion parameters for \ce{D2CS}. In addition, the \ce{\textit{cis}-DCSD} and \ce{\textit{trans}-DCSD} conformers, previously reported by \citet{INOSTROZAPINO2020111273} were also calculated with F12-TcCR level of theory. The rotational constants exhibit significant distinctions among the various isotopologues when compared to both each other and the primary isotopologue singlet \ce{D2CS}. Even though the differences in the a-type spectra, identified mainly by \textit{B+C}, for the \textit{cis} and \textit{trans} isomers, are quite small, the frequency resolution of modern facilities will enable their clear differentiation in laboratory experiments and/or astronomical observations.

Table \ref{Doubly-Deuterated Thioformaldehyde Isotopologues} presents the harmonic frequencies and fundamental vibrational frequencies for the isotopologues of  \ce{H2CS} such as \ce{$^1A_1$ D2CS}, \ce{$^1A_1$ D2SC}, \ce{$^3A''$ D2SC}, \ce{\textit{cis}-DCSD}, and \ce{\textit{trans}-DCSD}. As is commonly observed in the case of single-isotopic substitution, deuteration leads to the most significant alterations in frequencies, causing a decrease in their values for each mode. Here, a doubly-deuterated species exposed a significant variation. For example $\nu_1$  of \ce{H2CS} has a frequency of 3021.6 $cm^{-1}$ shifted to 2286.9, 2908.9, and 2837.2 $cm^{-1}$ for \ce{$^1A_1$ D$_{2}$CS},\ce{\textit{cis}-DCSD} and \ce{\textit{trans}-DCSD}, respectively.  Appendix \ref{H$_{2}C$^{34}$S Isotopologues_rc} shown the previously reported rotational constants of D$_{2}$C$^{34}$S isotopologues at CCSD(T)-F12/cc-pVTZ-F12 for comparison purposes. 
As such, these computed data should aid in laboratory characterization and/or interstellar examination of these molecules with the James Webb Space Telescope especially since D$_2$CS is known to exist in the ISM and other astronomical regions.

\begingroup
%\label{rotational}
\begin{table*} [h!]
\centering
\caption{New \emph{ab initio} Rotational Constants for Doubly-Deuterated H$_{2}$CS Isotopologues at the F12-TcCR level of theory [MHz].}
\label{table:2} \centering
\begin{tabular}{l|r r r r r r}
\hline\hline
  & $^1A_1$ D$_{2}$CS & $^1A_1$ D$_{2}$SC & $^3A''$ D$_{2}$SC & $cis$-DCSD & $trans$-DCSD \\
$A_e$        & 147374.4  &118605.4& 81000.6  &  101429.7 &104294.5&   \\
$B_e$        &  14967.4   & 17650.3  &  14965.9&  16945.5 & 16967.7  & \\
$C_e$        &  13587.5   &  15364.1&  14416.7 & 14519.8 & 14593.6&     \\
$A_0$        &146621.0  &  116161.7 & 80794.2& 103120.0 & 106332.6&    \\
$B_0$        & 14911.3    &  17557.6  & 14844.4  &17159.1&17170.9 &     \\
$C_0$        &13501.6     & 15250.8  &14282.7&  14668.6& 14744.1&     \\
$A_1$        & 145716.8 & 114302.3  & 80212.1 &105873.2 & 109273.8 &    \\
$B_1$        &14875.1 & 17527.5  &  14939.0 & 17207.1& 17203.3  &    \\
$C_1$        & 13471.5 & 15204.0  &  14380.8 & 14761.6 & 14827.5 &   \\
$A_2$        & 144659.8 & 114692.4& 79918.3 & 106459.0 &109709.1&  \\
$B_2$        & 14891.4 & 17512.8 & 14940.6 & 17153.9 & 17166.5   &     \\
$C_2$        &13465.9  & 15212.4  &14359.4& 14736.5& 14810.6   &    \\
$A_3$        & 147355.6 & 116291.9& 80239.0 & 103531.8 & 106727.3 &    \\
$B_3$        & 14893.7    &17445.9&  14917.5  &  17418.6 &17430.9&     \\
$C_3$        & 13437.3&15128.4&14316.0 & 14856.0& 14929.2 &     \\
$A_4$        &  146917.9  &116689.5&81358.2&102746.5 & 106501.6   &    \\
$B_4$        &   14889.3  &17554.1& 14619.1&  17239.8 &17251.8  &     \\
$C_4$        &  13467.9& 15195.5 &14097.7 & 14647.2 & 14727.5    &    \\
$A_5$        & 145243.7 &120482.3& 81276.7 &101471.2 & 107947.0    &    \\
$B_5$        & 14870.5 & 17611.5 & 14601.6 &17147.7 & 17173.4   &    \\
$C_5$        &13518.5      &  15200.1   &14101.9&  14666.5 & 14752.6  &    \\
$A_6$        &  148325.7   & 109624.9  & 81348.0  & 102018.8 & 101913.2  &   \\
$B_6$        &  14935.2     &  17507.9 &14804.9& 17214.5 &  17205.2   &    \\
$C_6$        &  13476.5    &  15338.2   &  14172.9   &14641.7&  14718.8  &    \\
$D_J$ (kHz)  &  12.694     &  16.290  &  44.744 & 20.596& 19.378   &    \\
$D_{JK}$       &0.279      &0.812&  0.491 & 0.213& 0.249    &      \\
$D_K$        & 5.402&4.117&1.061&2.148&  1.942 &     \\
$d_1$ (kHz)  & - 1.328    &-2.768&   -1.496  &  -3.269 & -2.879    &     \\
$d_2$ (kHz)  &-0.260    & -1.319   & -0.277 &  -0.621 &   -0.540   &     \\
$H_J$ (mHz)  &  1.369    &  -41.289   &  -320.953 & 133.294 & 137.559  &    \\
$H_{JK}$ (Hz)  &  0.863 & 4.830 & 0.180& 1.944   & 2.715    &     \\
$H_{KJ}$ (Hz)  &  -4.787  &  -92.073   &47.723&  -2.374 &  1.588   &    \\
$H_K$ (kHz)  &  0.660   &  0.577  & 0.110   &  0.055 & 0.147  &      \\
$h_1$ (mHz)  &   2.164  &-5.570 & -25.124  & 34.521 & 31.990      &      \\
$h_2$ (mHz)  &  2.318  & 19.712  & 5.794 & 11.223  & 10.208     &     \\
$h_3$ (mHz)  &   0.568  & 8.621  &  0.901 & 1.951 & 1.474      &    \\
$\kappa$     &  -0.97882  & -0.95428  & -0.98311  & -0.94369 & -0.94701     &    \\
$\mu_y$   &   &   &  & 2.16 &  0.59   &    \\
$\mu_z$    &   &  &   &  1.56  & 1.76  &  \\
$\mu$   (D)  &   &  &   &2.66 &    1.85&    \\

\hline
\end{tabular}
\label{Doubly-Deuterated Thioformaldehyde Isotopologues_rc}
\end{table*}
\endgroup

\begingroup
\begin{table*}[h]
\centering
\caption{ Harmonic and fundamental frequencies [cm$^{-1}$] of Doubly-Deuterated H$_{2}$CS Isotopologues at F12-TcCR level of theory.}

\begin{tabular}{l|r r r r r r}
\hline\hline

  & D$_{2}$CS & $^1A_1$ D$_{2}$SC & $^3A''$ D$_{2}$SC & $cis$-DCSD & $trans$-DCSD &  \\
$\omega_1$   &  2371.9  &   1630.7   &  1801.1  & 2250.7   & 2195.1   &    \\
$\omega_2$    &  2238.1  & 1582.6     &  1790.3  &  1744.4  & 1869.7   &    \\
$\omega_3$    &  1198.2  &  1068.8    &  848.8  &  921.6  &  927.2  &    \\
$\omega_4$    & 952.5   &   937.0   &  542.2  &  847.7  &  880.2  &    \\
$\omega_5$    &  791.8  &  452.2    &  495.3  & 701.6   & 718.3   &    \\
$\omega_6$    &  766.5  &  154.3    &  434.4  & 585.2   & 663.6   &    \\
$\nu_1$      &  2286.9  &  1527.5    &  1712.9  & 2173.3   &  2117.3  &    \\
$\nu_2$       &  2163.0  & 1469.7     & 1704.0   &  1669.7  &  1803.2  &    \\
$\nu_3$      &   1176.4 & 1044.9     & 823.8   &  908.4  &  908.2  &    \\
$\nu_4$       &  939.2  & 915.8     & 520.4   &  827.5  & 859.9   &    \\
$\nu_5$      & 783.9  &  450.8 & 474.4   &  689.1  & 709.5   &    \\
$\nu_6$      &  758.0 &  186.4 & 421.6   &  574.7  & 656.3   &    \\
Zero-Point    & 4123.4 &2880.9& 2915.7& 3491.8   & 3592.9   &    \\
\hline
\end{tabular}
\label{Doubly-Deuterated Thioformaldehyde Isotopologues}
\end{table*}
\endgroup

\newpage

\section{Conclusions}

In this work, accurate spectroscopic parameters for isotopologues of \ce{H2CS} are determined. The doubly-deuterated \ce{D2CS} isomers of thioformaldehyde are analyzed, providing laboratory and theoretical updates for this species. The new laboratory data allow for highly accurate frequency predictions well into the THz region to be used by the radio astronomical community. The agreement of the molecular parameters derived from this new experimental analysis and the new \textit{ab initio} calculations for {\dcs} is quite remarkable. The accuracy of the \textit{B} and \textit{C} rotational constants is below 0.1\% while the \textit{A} constant is only slightly higher. All the quartic and sextic parameters, except $d_2$ and the $h_1$-$h_3$ minor, sextic distortion constants, are predicted with an accuracy of a few percent from the experimental values. 

Since the identification of new sulfur-containing molecules can offer valuable insights into sulfur depletion, particularly in the case of the \ce{H2CS} isomers and apart from the widely recognized thioformaldehyde in astrochemistry, \ce{\textit{cis}-HCSH} stands out as the most likely candidate for detection. This likelihood arises from its significant dipole moment and notable vibrational intensities \citep{INOSTROZAPINO2020111273}. Considering that, we extended the work for these isomers but focused on double-deuterated species \ce{D2,C,S}. Thus, new sulfur-bearing molecules, such as \emph{cis}/\emph{trans}-DCSD isomers of \ce{D2CS}, were studied. Even though these isotopologues are less abundant than the corresponding H-analog, those species are prime candidates for astronomical observations. 

Further research is imperative to unravel the potentially hidden sulfur deposit within the ISM.

We demonstrated that using advanced quantum chemical methods in the computation of QFFs in conjunction with VPT2 through SPECTRO allows for the accurate prediction of the data constants necessary for obtaining the rotational spectrum of various molecules and their respective isotopologues. Thus, Dr.~Timothy J.~Lee laid the foundational theoretical principles that now yield results supporting spectroscopic molecular detections based on \emph{ab initio} methods.

\section{Acknowledgements}
We dedicate this article to the memory of Tim Lee, who passed away on November 3, 2022. His contributions have played a pivotal role in advancing the field of quantum chemistry as a whole and, especially, in the realm of astrochemistry. His absence will be deeply felt, and his legacy will continue to inspire and guide future generations.  

RCF acknowledges support from the University of Mississippi's College of Liberal Arts, the Mississippi Center for Supercomputing Research funded in part by NSF Grant OIA-1757220, and from NASA Grant 22-A22ISFM-0009.
NI gratefully acknowledges support of Vicerector\'{i}a de Investigaci\'{o}n y Postgrado (VRIP) and  PCI-ANID Grant REDES190113. 

%%%%%%%%%%%%%%%%%%%% REFERENCES %%%%%%%%%%%%%%%%%%

\bibliography{References} % your references Yourfile.bib

%%%%%%%%%%%%%%%%%%%%%%%%%%%%%%%%%%%%%%%%%%%%%%%%%%

\appendix
\section{Previous lower accurate QFF calculations}

\begingroup
\begin{table*}[h!]
\centering
\caption{Previously reported D$_{2}$C$^{34}$S Isotopologues CCSD(T)-F12/cc-pVTZ-F12 Rotational Constants [MHz].}

\begin{tabular}{l|r r r r r r}
\hline\hline

  & D$_{2}$C$^{34}$S & $^1A_1$ D$_{2}$$^{34}$SC & $^3A''$ D$_{2}$$^{34}$SC & $cis$-DC$^{34}$SD & $trans$-DC$^{34}$SD & $^3A$ DC$^{34}$SD\\
$A_e$        &  146983.0  &  118448.7  &  80605.9   &  100894.0  &  103538.8  &  117876.5 \\
$B_e$        &  14612.8   &  17439.6   &  14730.1   &  16608.9   &  16637.7   &  14451.1  \\
$C_e$        &  13291.4   &  15201.5   &  14197.9   &  14261.2   &  14334.3   &  13562.8  \\
$A_0$        &  146232.7  &  116076.9  &  80412.1   &  100568.7  &  102901.8  &  118252.2 \\
$B_0$        &  14557.5   &  17347.5   &  14611.1   &  16516.7   &  16551.3   &  14402.1  \\
$C_0$        &  13207.6   &  15089.7   &  14066.9   &  14145.5   &  14219.6   &  13475.0  \\
$A_1$        &  145330.5  &  114258.1  &  79829.7   &  99470.3   &  101715.4  &  117237.9 \\
$B_1$        &  14522.5   &  17319.4   &  14704.1   &  16478.9   &  16530.9   &  14350.0  \\
$C_1$        &  13178.5   &  15044.2   &  14163.9   &  14097.0   &  14183.5   &  13440.9  \\
$A_2$        &  144276.1  &  114645.7  &  79545.7   &  99632.0   &  101740.7  &  116249.4 \\
$B_2$        &  14538.0   &  17305.1   &  14705.1   &  16528.3   &  16564.6   &  14425.2  \\
$C_2$        &  13173.0   &  15054.7   &  14142.6   &  14138.5   &  14210.6   &  13473.7  \\
$A_3$        &  146994.8  &  116203.7  &  79861.5   &  100294.2  &  102814.3  &  118438.7 \\
$B_3$        &  14543.4   &  17241.4   &  14681.7   &  16393.5   &  16429.4   &  14299.1  \\
$C_3$        &  13146.8   &  14970.5   &  14098.8   &  14037.8   &  14112.4   &  13393.2  \\
$A_4$        &  146498.1  &  116605.9  &  81031.8   &  101489.0  &  104411.9  &  122457.0 \\
$B_4$        &  14533.5   &  17339.4   &  14396.8   &  16548.2   &  16572.7   &  14403.4  \\
$C_4$        &  13173.5   &  15033.3   &  13891.0   &  14116.6   &  14186.1   &  13473.4  \\
$A_5$        &  144859.5  &  120457.9  &  80857.9   &  99856.6   &  105262.2  &  119240.7 \\
$B_5$        &  14517.3   &  17399.7   &  14368.3   &  16484.4   &  16502.8   &  14421.5  \\
$C_5$        &  13223.7   &  15039.8   &  13882.8   &  14146.7   &  14210.6   &  13450.7  \\
$A_6$        &  147936.6  &  109546.4  &  80958.5   &  102020.0  &  100192.4  &  116640.9 \\
$B_6$        &  14579.7   &  17295.7   &  14572.3   &  16482.1   &  16534.2   &  14417.3  \\
$C_6$        &  13183.1   &  15172.0   &  13961.2   &  14105.3   &  14185.2   &  13440.2  \\
$D_J$ (kHz)  &  11.669    &  16.026    &  43.167    &  18.717    &  18.835    &  17.626   \\
$D_{JK}$       &  0.270     &  0.799     &  0.483     &  0.217     &  0.244     &  0.355    \\
$D_K$        &  5.387     &  4.105     &  1.063     &  2.115     &  1.898     &  4.171    \\
$d_1$ (kHz)  &  -1.245    &  -2.681    &  -1.451    &  -3.098    &  -2.756    &  -0.787   \\
$d_2$ (kHz)  &  -0.239    &  -1.267    &  -0.263    &  -0.584    &  -0.513    &  0.367    \\
$H_J$ (mHz)  &  1.344     &  -40.062   &  -311.924  &  -4.079    &  -4.100    &  14.128   \\
$H_{JK}$ (Hz)  &  0.789     &  4.662     &  0.163     &  0.679     &  0.013     &  -1.203   \\
$H_{KJ}$ (Hz)  &  -4.687    &  -90.705   &  47.375    &  -5.982    &  11.473    &  -37.485  \\
$H_K$ (kHz)  &  0.656     &  0.590     &  0.110     &  0.284     &  0.151     &  0.690    \\
$h_1$ (mHz)  &  1.991     &  -5.439    &  -23.781   &  5.210     &  3.916     &  2.281    \\
$h_2$ (mHz)  &  2.066     &  18.644    &  5.464     &  6.331     &  4.107     &  -5.100   \\
$h_3$ (mHz)  &  0.501     &  8.146     &  0.835     &  1.614     &  1.253     &  -2.841   \\

\hline
\end{tabular}
\label{H$_{2}C$^{34}$S Isotopologues_rc}
\end{table*}
\endgroup

\end{document}